\documentclass[preprint,showpacs,amsmath,amssymb]{revtex4}
\usepackage{changes}
\usepackage{amsfonts}
\usepackage{amsmath,graphicx,color,epsfig}

\newcommand{\be}{\begin{equation}}
\newcommand{\ee}{\end{equation}}
\newcommand{\bea}{\begin{eqnarray}}
\newcommand{\eea}{\end{eqnarray}}
\newcommand{\ba}{\begin{array}}
\newcommand{\ea}{\end{array}}
\newcommand{\bi}{\begin{itemize}}
\newcommand{\ei}{\end{itemize}}

\newcommand{\mca}{{\mathcal A}}
\newcommand{\mcb}{{\mathcal B}}

\newcommand{\mcf}{{\mathcal F}}

\newcommand{\mcm}{{\mathcal M}}
\newcommand{\mcn}{{\mathcal N}}

\renewcommand{\vec}[1]{\mbox{\boldmath $#1 \!\!$ \unboldmath}}

\newcommand{\g}{\gamma}
\newcommand{\nslash}{\kern 0.2 em n\kern -0.50em /}
\newcommand{\kslash}{\kern 0.2 em k\kern -0.45em /}
\newcommand{\qslash}{\kern 0.2 em q\kern -0.45em /}
\newcommand{\pslash}{\kern 0.2 em p\kern -0.50em /}
\newcommand{\rslash}{\kern 0.2 em r\kern -0.50em /}
\newcommand{\sslash}{\kern 0.2 em s\kern -0.50em /}
\newcommand{\Sslash}{\kern 0.2 em S\kern -0.50em /}
\newcommand{\Pslash}{\kern 0.2 em P\kern -0.50em /}
\newcommand{\Dslash}{\kern 0.2 em D\kern -0.65em /\kern 0.15em}
\newcommand{\lf}{\left}
\newcommand{\rg}{\right}

\newcommand{\de}{d}                    

\newcommand{\half}{ {\textstyle\frac{1}{2}} }

\begin{document}
\title{Single-spin asymmetries in SIDIS induced by anomalous quark-gluon and quark-photon couplings}

\author{Xu Cao$^{1}$ \footnote{caoxu@impcas.ac.cn}}
\author{Nikolai Korchagin$^{1,2}$ \footnote{korchagin@impcas.ac.cn}}
\author{Nikolai Kochelev$^{1,2}$ \footnote{kochelev@theor.jinr.ru}}
\author{Pengming Zhang$^{1}$ \footnote{zhpm@impcas.ac.cn}}
\affiliation{$^1$CAS Key Laboratory of High Precision Nuclear Spectroscopy and Center for Nuclear Matter Science, Institute of Modern Physics, Chinese Academy of Sciences, Lanzhou 730000, People Republic of China\\
$^2$Bogoliubov Laboratory of Theoretical Physics, Institute for Nuclear Research, Dubna, Moscow Region 141980, Russia}

\begin{abstract}
  Abstract: We calculate the contribution of the non-perturbative Pauli couplings in the quark-photon and quark-gluon vertices to the single-spin asymmetries (SSAs) in semi-inclusive deep inelastic scattering (SIDIS). We describe the nucleon using the spectator model with scalar and axial-vector diquarks. Both of the new couplings can cause the helicity flip of the struck quark. The helicity flip and the rescattering induced by  the non-perturbative gluon exchange between the struck quark and diquark lead to the SSAs. Their azimuthal dependencies are the same as that usually ascribed to the Collins and Sivers effects. Our numerical results, based on the instanton model for the QCD vacuum, show that the non-perturbative quark-gluon  and quark-photon interactions have a strong influence on these asymmetries.


\end{abstract}
\maketitle

%

\section{Introduction} \label{sec:intro}

One of the long-standing problems in the strong interaction theory is to understand the origin of spin effects in hadronic physics
within Quantum Chromodynamics (QCD). To achieve this goal, a promising route is to investigate the mechanisms that are responsible
for the large single-spin asymmetries (SSAs) in high energy hadronic reactions and in semi-inclusive deep-inelastic
scattering (SIDIS)~\cite{Barone:2010zz,Aschenauer:2015ndk,Boer:2011fh}.
Two main mechanisms based on QCD factorization are usually  considered. One of them is
related to the spin-dependent Sivers distribution function \cite{Sivers:1989cc,Sivers:1990fh} and the other comes from
the spin-dependent Collins fragmentation function  ~\cite{Collins:1992kk,Collins:2002kn}.
These functions should be either calculated within some nonperturbative approach based on QCD
\cite{Yuan:2003wk,Cherednikov:2006zn,Courtoy:2008vi} or  be extracted
directly from experiments  \cite{Vogelsang:2005cs,Martin:2017yms,Avakian:2016rst,Airapetian:2013bim,Anselmino:2013rya}.
Another way to produce a non-zero SSA is through final or initial interactions between quarks
by perturbative gluon exchange
~\cite{Brodsky:2002cx}. There is a relation between the two mechanisms  \cite{Ji:2002aa}.
Recently the twist-3 mechanism, based on quark-gluon correlations \cite{Efremov:1981sh,Efremov:1984ip,Qiu:1991wg,Qiu:1998ia}, was revisited
\cite{Kang:2010zzb,Kouvaris:2006zy,Kanazawa:2015ajw,Metz:2012ct,Mao:2014aoa}
to solve the problem  of the ``sign-mismatch" between the Sivers function
extracted from the inclusive production of pions in proton-proton collisions and SIDIS  \cite{Kang:2011hk,Kang:2012xf}.
Most of the approaches to calculate SSAs are based on the factorization assumption (see reviews \cite{Pitonyak:2016hqh,Metz:2016swz}).
However, a proof of the validity of such an assumption is still lacking \cite{Rogers:2010dm,Rogers:2013zha}.
Furthermore, the explicit mechanism for the breakdown of transverse-momentum-dependent (TMD) factorization
was suggested in \cite{Kochelev:2013zoa} (see also \cite{Kochelev:2015pha}). This mechanism is based on the existence of a small
size strong gluonic fluctuation in the QCD vacuum called instantons (see reviews \cite{Schafer:1996wv,Diakonov:2002fq}).
These nontrivial topological solutions of QCD produce a
very large anomalous quark chromomagnatic moment (AQCM) which flips the quark helicity \cite{Kochelev:1996pv}.
This quark helicity flip is one of the important ingredients to generate SSAs.  Indeed,  it was demonstrated in Ref. \cite{Kochelev:2013zoa}
that the quark-gluon vertex induced by AQCM leads to a large
SSA in quark-quark scattering (see also \cite{Ostrovsky:2004pd,Qian:2015wyq} and references therein). Additionally, the instantons induce an
anomalous quark magnetic moment~\cite{Zhang:2017zpi} and, therefore, produce a nontrivial quark-photon vertex which flips the quark helicity and gives rise to SSA in SIDIS. It should be mentioned that the first qualitative discussion about possible effects of the
Pauli-type soft quark-gluon interaction on SSAs in SIDIS was in Ref.\cite{Hoyer:2005ev}.
The evidence for the existence of  the smaller scale in QCD, compared to the
confinement scale, was discussed in Refs.~\cite{Dorokhov:1993fc,Kopeliovich:2007pq,Schweitzer:2012hh,Kochelev:2015pqd}.

In this paper we calculate SSAs in SIDIS using the instanton model for the helicity flip  in the quark-gluon and quark-photon
vertices \cite{Kochelev:1996pv,Zhang:2017zpi}. To describe the proton we use the spectator model \cite{Bacchetta:2003rz,Bacchetta:2008af} with
the non-perturbative final state interaction between the struck quark and both the scalar and axial-vector diquarks.
We report the full set of helicity amplitudes, taking into account this novel helicity non-conserving mechanisms and study
the angular dependence of SSA observables.

\section{Analytical results in spectator models} \label{sec:resul}

\begin{figure}
\begin{center}
{\includegraphics*[width=12.cm]{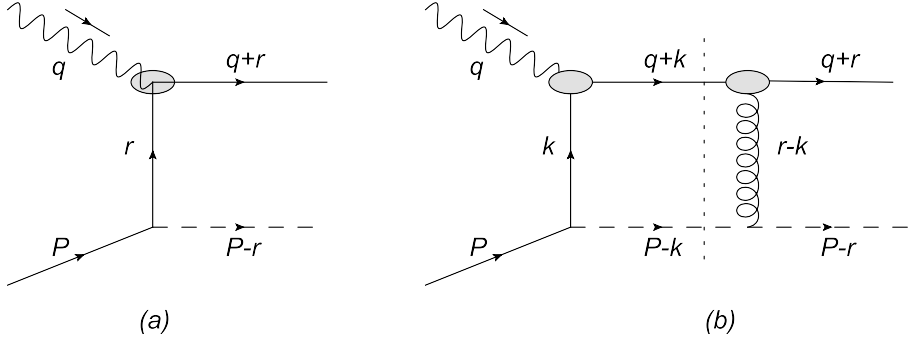}}
\caption{The interference of the tree diagram (a) with the one-loop diagram (b) provides the SSA. Amplitudes $\mca$ and $\mcb$ correspond the left and right diagrams, respectively. The quark-diquark rescattering amplitude $\mcm$ corresponds to the right part of the loop diagram. The blobs represent the interaction vertices with both Dirac and Pauli couplings.
\label{fig:SSAFeynman}}
\end{center}
\end{figure}
In our model SSA comes from the interference of the tree and one-loop diagrams
presented in FIG.~\ref{fig:SSAFeynman}. The general quark-gluon vertex for the interaction between the struck quark and the exchanged gluon
has the following form~\cite{Hoyer:2005ev}
\be \label{eq:vertexgqq}
V^{g,a}_\mu = t^a g_s \lf( \mcf_D^g \g^{\mu} - \frac{\mcf_P^g}{2m_q} \sigma^{\mu\nu} (r_\nu-k_\nu) \rg),
\ee
where $t^a$ are the Gell-Mann colour matrices, $g_s$ is the strong coupling constant, $m_q$ is the constituent quark mass, $\sigma_{\mu\nu}=[\gamma_{\mu},\gamma_{\nu}]/2$ and $r-k$ is the gluon momentum. We also consider the general vertex
for the interaction of the struck quark with the virtual photon~\cite{Zhang:2017zpi}
\be \label{eq:vertexpqq}
V^\gamma_\mu=\mcf_D^\gamma \g^{\mu} - \frac{\mcf_P^\gamma}{2m_q} \sigma^{\mu\nu} q_\nu,
\ee
where $q$ is the photon momentum, $\mcf_D^g (\mcf_D^\gamma) $ and $\mcf_P^g (\mcf_P^\gamma) $ are Dirac and Pauli form factors, which are functions of the corresponding gluon or photon momentum, respectively.

The interaction vertices of the gluon with the scalar and axial-vector diquarks are \cite{Bacchetta:2003rz,Bacchetta:2008af}
\bea
\Pi^s_{\mu} &=& i e_s (2 P - k - r)_{\mu}F^s_{\rm di}, \\
\Pi^a_{\mu,\alpha\beta} &=& - i e_a \lf[ (2 P - k - r)_{\mu} g_{\alpha\beta} -(P -  k)_{\alpha} g_{\mu\beta} - (P - r)_{\beta} g_{\alpha\mu} \rg]F^a_{\rm di}, \nonumber
\eea
where $e_s$ and $e_a$ are the color charges of the scalar and axial-vector diquarks
and $F^{s,a}_{\rm{di}}$ are the corresponding diquark form factors. The nucleon-quark-diquark vertices are chosen to be
\bea
V_s &=& i g_P^s\bf 1,
\\ V_{\mu}^a &=& i \frac{g_P^a}{\sqrt{2}} \gamma_5 \gamma^\mu, \nonumber
\eea
where  $\bf 1$ is the unit matrix in the spinor space, $g_P^s$ and $g_P^a$ are the couplings in the proton-quark-scalar diquark vertex
and the proton-quark-axial-vector diquark vertex.
In general, these vertices depend on momenta of particles, i.e. they should include form factors (see \cite{Bacchetta:2008af} and references therein).
However, we limit ourselves to the simplest point-like case.

To calculate helicity amplitudes we will use the approach presented by Hoyer and Jarvinen  in \cite{Hoyer:2005ev}.
Following their method, we  choose a reference frame where the target proton is at rest and has the spin in the $y$ direction. The virtual photon momentum is along the $+z$ axis, i.e. the momenta of photon, proton and struck quarks are~\cite{Hoyer:2005ev}
\bea \label{eq:cmsystem}
 q &=& (q^+,q^-,0_\perp) \simeq  (2\nu,-x M,\vec{0}_\perp),
 \nonumber \\
 P &=& (M,M,\vec{0}_\perp), \\
 k &=& (xM,xM,\vec{k}_\perp = k_\perp e^{i\phi}),
 \nonumber \\
 r &=& (xM,xM,\vec{r}_\perp = r_\perp e^{i\psi}), \nonumber
\eea
where $\nu = Q^2/2xM$ is the photon energy, $M$ is the proton mass and $x$ is the Bjorken variable. The polarization vectors of the photon and the axial-vector diquark are defined as
\bea
\epsilon^\lambda(q) &=& \frac{1}{\sqrt{2}} \lf(0, 0, -\lambda, - i\rg),
\\
\epsilon_D(P-k,\lambda_a) &=& \frac{1}{\sqrt{2}} \lf(\frac{2(\lambda_a k_x + i k_y)}{(1-x) M}, 0, -\lambda_a, - i\rg). \nonumber
\eea
They satisfy the transversality condition, for instance $(P-k) \cdot \epsilon_D(P-k,\lambda_a) = 0$. The amplitude for the elastic quark-scalar diquark scattering process is $\mcm_{s,s'}$. It corresponds to the right-hand side of  FIG.~\ref{fig:SSAFeynman}(b). $s$ and $s'$ are helicities of the initial and final quarks, respectively. In the limit of $q^+ = Q^2/xM \to\infty$ at fixed $k$ and $r$, the helicity amplitudes are
\bea
\mcm_{+,+} &\simeq& F^s_{\rm di} \mcf_D^g \, \frac{e_qe_{s}}{(\vec{k}_\perp-\vec{r}_\perp)^2} \, 2 (1-x) M q^+,  \\
\mcm_{+,-} &\simeq& - F^s_{\rm di} \frac{\mcf_P^g}{2m_q} \, \frac{e_qe_{s}}{(\vec{k}_\perp-\vec{r}_\perp)^2} \, 2 (1-x) M q^+ \left(k_\perp e^{+i\phi}-r_\perp e^{+i\psi}\right) \nonumber,
\eea
where $e_q$ is the color charge of the struck quark and $F^s_{\rm di}$ is the scalar diquark form factor.
For the Born diagram presented in FIG.~\ref{fig:SSAFeynman}(a), the helicity amplitude is $\mca^{\lambda}_{s,s'}$, where $\lambda$ is the photon helicity. The amplitudes are
\bea
\mca^{\lambda}_{+,+} &\simeq& Q_qg_P^s \sqrt{2M q^+} \, \frac{1-x}{\vec{r}_\perp^2+B_R^2(m_q^2)} \, \times \nonumber \\
 ~&~&\lf[ \lf(\mcf_D^\gamma - \frac{\mcf_P^\gamma}{2m_q} D_Q\rg) \, r_\perp e^{+i\psi}\, \delta_{\lambda,+1} + \frac{\mcf_P^\gamma}{2m_q}\, D_R \, r_\perp e^{-i\psi} \,\delta_{\lambda,-1} \rg], \\
\mca^{\lambda}_{+,-} &\simeq&  Q_qg_P^s \sqrt{2M q^+} \, \frac{1-x}{\vec{r}_\perp^2+B_R^2(m_q^2)} \,\lf[ -\frac{\mcf_P^\gamma}{2m_q} \vec{r}_\perp^2 e^{2i\psi} \, \delta_{\lambda,+1} + \lf( \mcf_D^\gamma+ \frac{\mcf_P^\gamma}{2m_q} D_R \rg) D_R \, \delta_{\lambda,-1} \rg], \nonumber
\eea
where $D_Q = x M-m_q$, $D_R = x M+m_q$, $B_R^2(m_q^2) = (1-x)m_q^2 +x m_D^2 - x(1-x) M^2$ and $Q_q$ is the electric charge of the struck quark.

 For the axial-vector diquark case the amplitude for the quark-diquark scattering is $\mcm_{s,s'}^{\lambda_a,\lambda_a'}$. Here $\lambda_a$ and $\lambda_a'$ are helicities of the initial and final diquark, respectively. Amplitudes are
\bea
\mcm_{+,+}^{\lambda_a,\lambda_a'} &\simeq& F^a_{\rm di} \mcf_D^g \, \frac{e_qe_{a}}{(\vec{k}_\perp-\vec{r}_\perp)^2} \, 2 (1-x) M q^+ \, \delta_{\lambda_a,\lambda_a'}, \\
\mcm_{+,-}^{\lambda_a,\lambda_a'} &\simeq& - F^a_{\rm di} \frac{\mcf_P^g}{2m_q} \, \frac{e_qe_{a}}{(\vec{k}_\perp-\vec{r}_\perp)^2} \, 2 (1-x) M q^+ \left(k_\perp e^{+i\phi}-r_\perp e^{+i\psi}\right) \, \delta_{\lambda_a,\lambda_a'}, \nonumber
\eea
where $F^a_{\rm di}$ is the axial-vector diquark form factor. For the diagram on FIG.~\ref{fig:SSAFeynman}(a) with the axial-vector diquark the amplitude is $\mca^{\lambda,\lambda_a}_{s,s'}$, where $\lambda$ and $\lambda_a$ are photon and diquark helicities, respectively. The full set of amplitudes is
\bea
\mca^{+,+}_{+,+} &\simeq& -  Q_qg_P^a  \sqrt{2M q^+} \, \frac{1-x}{\vec{r}_\perp^2+B_R^2(m_q^2)} \, \lf[ \lf( \mcf_D^\gamma + \frac{\mcf_P^\gamma}{2m_q} D_R \rg) D_R + \frac{\mcf_P^\gamma}{2m_q} \frac{x}{1-x} \,\vec{r}_\perp^2\rg], \nonumber \\
\mca^{+,-}_{+,+} &\simeq&  Q_qg_P^a  \sqrt{2M q^+} \, \frac{1-x}{\vec{r}_\perp^2+B_R^2(m_q^2)} \frac{\mcf_P^\gamma}{2m_q} \frac{x}{1-x} \,\vec{r}_\perp^2 e^{+2i\psi}, \nonumber \\
\mca^{-,+}_{+,+} &\simeq& Q_qg_P^a \sqrt{2M q^+} \, \frac{1-x}{\vec{r}_\perp^2+B_R^2(m_q^2)} \, \frac{\mcf_P^\gamma}{2m_q} \frac{1}{1-x} \,\vec{r}_\perp^2 e^{-2i\psi}, \nonumber \\
\mca^{-,-}_{+,+} &\simeq& -  Q_qg_P^a \sqrt{2M q^+} \, \frac{1-x}{\vec{r}_\perp^2+B_R^2(m_q^2)} \,\frac{\mcf_P^\gamma}{2m_q} \frac{x}{1-x} \, \vec{r}_\perp^2,  \\
\mca^{+,+}_{+,-} &\simeq&  Q_qg_P^a \sqrt{2M q^+} \, \frac{1-x}{\vec{r}_\perp^2+B_R^2(m_q^2)} \, \frac{\mcf_P^\gamma}{2m_q} \, D_R \, r_\perp e^{+i\psi}, \nonumber\\
\mca^{+,+}_{-,+} &\simeq& -  Q_qg_P^a \sqrt{2M q^+} \, \frac{1-x}{\vec{r}_\perp^2+B_R^2(m_q^2)} \lf(\mcf_D^\gamma + \mcf_P^\gamma \rg) \frac{x}{1-x} r_\perp e^{-i\psi}, \nonumber \\
\mca^{+,-}_{-,+} &\simeq&  Q_qg_P^a  \sqrt{2M q^+} \, \frac{1-x}{\vec{r}_\perp^2+B_R^2(m_q^2)} \, \lf[ \mcf_D^\gamma  + \frac{\mcf_P^\gamma}{2m_q} \lf( x\,D_R - D_Q \rg) \rg] \frac{r_\perp e^{+i\psi}}{1-x}, \nonumber \\
\mca^{+,-}_{+,-} &\simeq& 0. \nonumber
\eea
The remaining helicity amplitudes are obtained through the relations
\be
\mcm_{s,s'}^{(\lambda_a,\lambda_a')} = (-1)^{s-s'} \lf(\mcm_{-s,-s'}^{(-\lambda_a,-\lambda_a')}\rg)^* \quad\text{and}\quad \mca_{s,s'}^{\lambda(,\lambda_a)} = -(-1)^{s-s'} \left(\mca_{-s,-s'}^{-\lambda(,-\lambda_a)}\right)^*. \nonumber
\ee
Only the discontinuity (absorptive part) of the loop amplitude $\mcb$ in FIG.~\ref{fig:SSAFeynman}(b) contributes to the asymmetry $A_N$.
According to the Cutkosky rules the discontinuity is given by a convolution of the amplitudes $\mca^{\lambda(,\lambda_a')}_{s,s''}$ and $\mcm_{s'',s'}^{(\lambda_a',\lambda_a)}$~\cite{Hoyer:2005ev},
\bea
\textrm{Disc}\, \mcb^{\lambda(,\lambda_a)}_{s,s'} &=& i \int \frac{\de^4 k}{(2\pi)^4} {\mskip 2.5mu} 2\pi {\mskip 2.5mu}\delta\lf( (k+q)^2 - m_q^2 \rg) {\mskip 2.5mu} 2\pi {\mskip 2.5mu} \delta\lf( (P-k)^2 - M_D^2 \rg) \sum_{s'',\lambda_a'} \mca^{\lambda(,\lambda_a')}_{s,s''} \mcm_{s'',s'}^{(\lambda_a',\lambda_a)} \nonumber \\ &=& \frac{i }{2(1-x)M q^+} \int \frac{\de^2 \vec{k}_{\bot}}{(2\pi)^2} {\mskip 2.5mu} \sum_{\lambda_a'} \lf[\mca^{\lambda(,\lambda_a')}_{s,-s'} \mcm_{-s',s'}^{(\lambda_a',\lambda_a)} + \mca^{\lambda(,\lambda_a')}_{s,s'} \mcm_{s',s'}^{(\lambda_a',\lambda_a)} \rg], \label{disc}
\eea
which satisfies
\be
\textrm{Disc}\, \mcb^{\lambda(,\lambda_a)}_{s,s'} = (-1)^{s-s'} \left(\textrm{Disc}\, \mcb_{-s,-s'}^{-\lambda(,-\lambda_a)} \right)^*.
\nonumber\ee
In the scalar diquark model considered in Ref.~\cite{Hoyer:2005ev} only the first term in Eq.~(\ref{disc}) was presented  due to the fact that
$\mcm_{s,s} \simeq 0$ there.   In contrast, $\mcm_{s,s} \propto \mcf_D^g$ in our calculation.
The spin asymmetry for a target polarized in the transverse $y$ direction is defined as
\bea
\mcn A_N &\simeq& \frac{2}{Q^4} \sum_{\lambda(,\lambda_a),\lambda'(,\lambda_a'),s'} \mathrm{Im}\lf\{ L^{\lambda,\lambda'} \lf[ \mca^{\lambda(,\lambda_a)}_{+,s'} + \half \textrm{Disc}\mcb^{\lambda(,\lambda_a)}_{+,s'} \rg] \lf[ \mca^{\lambda'(,\lambda_a')}_{-,s'} + \half \textrm{Disc}\mcb^{\lambda'(,\lambda_a')}_{-,s'} \rg]^* \rg\} \qquad \nonumber
\\
\qquad
&=& \frac{4 e^2}{y^2 Q^2} \mathrm{Im} \sum_{\lambda(,\lambda_a)} \left[1+(1-y)^2\right]\lf\{ \mca^{\lambda(,\lambda_a)}_{+,-} \textrm{Disc}\mcb^{-\lambda(,-\lambda_a)}_{+,+} - \mca^{\lambda(,\lambda_a)}_{+,+} \textrm{Disc}\mcb^{-\lambda(,-\lambda_a)}_{+,-}  \rg\} \nonumber
\\
&& \quad\quad\quad\quad - 2(1-y) e^{-2i\lambda\tau} \lf\{ \mca^{\lambda(,\lambda_a)}_{+,-} \textrm{Disc}\mcb^{\lambda(,-\lambda_a)}_{+,+} - \mca^{\lambda(,\lambda_a)}_{+,+} \textrm{Disc}\mcb^{\lambda(,-\lambda_a)}_{+,-} \rg\}, \label{eq:ANdef}
\eea
with the leptonic tensor in the helicity basis
\be \label{eq:Lexpr}
L^{\lambda,\lambda'} = \frac{2e^2 Q^2}{y^2} \left\{\left[1+(1-y)^2\right]\delta_{\lambda,\lambda'} - 2(1-y) e^{-2i\lambda\tau}\delta_{\lambda,-\lambda'} \right\}.
\ee
Here $\tau$ is the azimuthal angle of the lepton $\textbf{\emph{l}}_{1\perp}=\textbf{\emph{l}}_{2\perp}=l_\perp e^{i\tau}$ and $y=P \cdot q / P\cdot l_1$ is the fraction of the beam energy carried by the virtual photon. Note that the $\delta_{\lambda,-\lambda'}$ term in $L^{\lambda,\lambda'}$ generates the asymmetry in the third line of Eq.~(\ref{eq:ANdef}),
which was in the calculation by Hoyer and Jarvinen \cite{Hoyer:2005ev}.
The $\delta_{\lambda,\lambda'}$ term produces the second line of Eq.~(\ref{eq:ANdef}), and
leads to an asymmetry similar to the Sivers effect as will be demonstrated below. This asymmetry arises in our model because the amplitudes $\mca^{\lambda(,\lambda_a)}_{-s,s}$ and the helicity non-flip amplitudes $\mcm_{s,s}^{(\lambda_a',\lambda_a)}$ are generally non-zero. The normalization $\mcn$ is given by the amplitude at tree order in FIG.~\ref{fig:SSAFeynman}(a):
\bea
\mcn &=& \frac{1}{Q^4} \sum_{\lambda(,\lambda_a),\lambda'(,\lambda_a'),s,s'}L^{\lambda,\lambda'} \mca^{\lambda(,\lambda_a)}_{s,s'} \lf( \mca^{\lambda'(,\lambda_a')}_{s,s'} \rg)^* \nonumber \\ &=& \frac{8 Q_q^2 (g_P^{s,a})^2}{y^2 Q^2} M q^+ \lf(\frac{1-x}{\vec{r}_{\bot}^2 +B_R^2(m_q^2)}\rg)^2\left[1+(1-y)^2\right] \mcn^i_{+},
\eea
where $\mcn^i_{+}$ depends on a diquark model. For the scalar diquark, we have
\bea
\mcn_{+}^s &=& \sum_{\lambda} \lf\{ |\mca^{\lambda}_{+,+}|^2 + |\mca^{\lambda}_{+,-}|^2 \rg\} \nonumber \\ &\simeq& \lf(\mcf_D^\gamma - \frac{\mcf_P^\gamma}{2m_q} D_Q \rg)^2 \,\vec{r}_{\bot}^2 + \lf(\mcf_D^\gamma + \frac{\mcf_P^\gamma}{2m_q} D_R \rg)^2 \,D_R^2 + \lf(\frac{\mcf_P^\gamma}{2m_q}\rg)^2 \,\vec{r}_{\bot}^2 \,(\vec{r}_{\bot}^2+D_R^2), \quad
 \eea
which is in agreement with  the results of Hoyer and Jarvinen \cite{Hoyer:2005ev} in the limit  $\mcf_P^\gamma \rightarrow 0$. For the axial-vector diquark model, we obtain
\bea
\mcn_+^v &=& \sum_{\lambda,\lambda_a} \lf\{ |\mca^{\lambda,\lambda_a}_{+,+}|^2 + |\mca^{\lambda,\lambda_a}_{+,-}|^2 \rg\} \nonumber \\ &\simeq& \lf[ \lf( \mcf_D^\gamma + \frac{\mcf_P^\gamma}{2m_q} D_R \rg) D_R + \frac{\mcf_P^\gamma}{2m_q} \frac{x}{1-x} \,\vec{r}_\perp^2 \rg]^2 + \frac{\vec{r}_\perp^2}{(1-x)^2} \times
\nonumber \\
&& \Big[ \lf(\mcf_D^\gamma + \mcf_P^\gamma \rg)^2\,x^2 + \lf( \mcf_D^\gamma  + \frac{\mcf_P^\gamma}{2m_q} \lf( x\,D_R - D_Q \rg) \rg)^2 + \nonumber
\\
&& \lf(\frac{\mcf_P^\gamma}{2m_q} \rg)^2 \,\lf((1+2\,x^2)\,\vec{r}_\perp^2 + (1-x)^2\,D_R^2\rg) \Big].
\eea
It can be shown that the unpolarized distributions $\mcn_+^{s,a}$ agree with well known results of conventional diquark models
when $\mcf_P^\gamma$ vanishes.
Using  Eq.~(\ref{eq:ANdef}) we find that the asymmetry is
\bea \label{eq:ANscalc}
 A_N^{s,a} &\simeq & e_qe_{s,a} \dfrac{\vec{r}_\perp^2+B_R^2(m_q^2)}{\left[1+(1-y)^2\right] \mcn_{+}^{s,a}} \times
\nonumber \\
&& \Big\{ 2(1-y) \lf[ \mcf^+_{s,a} J_{s,a}^+ \cos(3\psi-2\tau) + \mcf^-_{s,a} J^-_{s,a} \cos(\psi-2\tau) \rg]  +
 \nonumber \\
 &&\left[1+(1-y)^2\right] \mcf^0_{s,a} J^0_{s,a} \cos\psi \Big\},
\eea
with the abbreviated notations $\mcf^n_{s,a}$ and $J_{s,a}^n$ are explained in the Appendix. The $\mcf^n_{s,a}$ functions are combinations of Dirac and Pauli couplings, and the loop functions $J_{s,a}^n$ are regularized by the form factors in the Dirac and Pauli couplings, which will be presented in Sec.~\ref{sec:formfactor}.

In the Trento convention \cite{Bacchetta:2004jz}, the angles $\phi_s$ and $\phi_h$ are used. The $\phi_s$ is defined as the angle between
the target spin direction and the lepton plane
\be \label{eq:Trentophi}
 \phi_s = \pi/2-\tau.
\ee
The $\phi_h$ is the angle between the hadron (or quark jet) and lepton planes
\be \label{eq:Trentopsi}
 \phi_h = \psi-\tau.
\ee
Using these coordinates, the asymmetry in the both models is
\be \label{eq:ANfinal}
A_N = \epsilon A_N^{\sin(3\phi_h-\phi_s)} \sin(3\phi_h-\phi_s) + \epsilon A_N^{\sin(\phi_h+\phi_s)} \sin(\phi_h+\phi_s) + A_N^{\sin(\phi_h-\phi_s)}
\sin(\phi_h-\phi_s),
\ee
with the depolarization factor $\epsilon = 2 (1-y)/ [1+(1-y)^2]$ and the definition:
\bea
A_N^{\sin(3\phi_h-\phi_s)} &=& - e_qe_{s,a} \frac{\vec{r}_\perp^2+B_R^2(m_q^2)} {\mcn_{+}^{s,a}} \mcf^+_{s,a} J_{s,a}^+
{\mathbf ,}
\\
A_N^{\sin(\phi_h+\phi_s)} &=& e_qe_{s,a} \frac{\vec{r}_\perp^2+B_R^2(m_q^2)} {\mcn_{+}^{s,a}} \mcf^-_{s,a} J_{s,a}^-
{\mathbf ,}
\\
A_N^{\sin(\phi_h-\phi_s)} &=& - e_qe_{s,a} \frac{\vec{r}_\perp^2+B_R^2(m_q^2)} {\mcn_{+}^{s,a}} \mcf^0_{s,a} J_{s,a}^0
{\mathbf .}
\eea
The second line in Eq.~(\ref{eq:ANdef}) gives the  asymmetry resulting from the relation $\cos\psi = - \sin(\phi_h-\phi_s)$
and the angular dependence of the asymmetry is identical to that which arises from the Sivers effect. The third line with $\lambda = +1$ in Eq.~(\ref{eq:ANdef}) gives
the asymmetry resulting from $\cos(3\psi-2\tau) = - \sin(3\phi_h-\phi_s)$ and when $\lambda = -1$ this gives
the asymmetry resulting from $\cos(\psi-2\tau) = \sin(\phi_h+\phi_s)$. Their angular dependence is the same as that from the Collins effect.
The physical origin of the two asymmetries is, however, quite different from the original Collins and Sivers asymmetry.
The helicity flip, which is the origin of all these asymmetries arises either from the tree level amplitudes
$\mca^{\lambda(,\lambda_a)}_{s,s'}$ or from the loop amplitudes $\textrm{Disc}\mcb^{\lambda(,\lambda_a)}_{s,s'}$.
As a result, the Pauli couplings  in both the quark-gluon and quark-photon vertices are able to generate the SSA.
In the axial diquark model the Pauli couplings of the quark-photon vertex is required to induce the $\sin(3\phi_h-\phi_s)$ asymmetry.
This is  because the factors $\mcf^+_{a,1}$, $\mcf^+_{a,2}$ and $\mcf^+_{a,3}$ are all proportional to $\mcf_P^\gamma$
since they are related to the $\mca^{+,-}_{+,+}$ amplitude in Eq.~(\ref{eq:ANdef}), which is also proportional to $\mcf_P^\gamma$.

\section{Form factors} \label{sec:formfactor}

For definiteness we will consider the case when the struck quark is a u-quark and the spectator is a (ud)-diquark.
The Pauli form factor ${\mcf_P^g}$ of the quark-gluon interaction was calculated using the instanton liquid model for
the QCD vacuum in Refs.~\cite{Kochelev:1996pv,Kochelev:2009rf,Kochelev:2013csa}. It is
\be \label{eq:AQCM}
\mcf_P^g((r-k)^2)  = \mu_a F^g_P((r-k)^2),
\ee
where
\be \label{eq:AQCM2}\mu_a = - \frac{3\pi (\rho_c m_q)^2}{4\alpha_s(\rho_c)}
\ee
and
\be
 F^g_P((r-k)^2)\approx e^{-(r-k)^2/\Lambda_q^2}
\ee
with $\Lambda_q = 2/\rho_c = 1.2$ GeV and $\rho_c=1/(600~{\rm MeV})$ is the instanton size~\cite{Zhang:2017zpi}.

We take the the diquark radius from the instanton liquid model \cite{Cristoforetti:2004kj} which leads to the diquark form factor
\be
 F^{s,a}_{\rm di}((r-k)^2)\approx e^{-(r-k)^2/\Lambda_{di}^2},
\ee
where $\Lambda_{di}\approx 0.7 $ GeV for both the scalar and axial-vector diquark.

The Pauli form factor in the quark-photon vertex from the non-perturbative contribution has been calculated within
the instanton  model \cite{Zhang:2017zpi} and it can be approximated well by
\be \label{eq:pauliq}
\mcf_P^\gamma(Q^2) = \frac{\mu_q}{1 + \rho_c Q^2/(4.7 m_q)}.
\ee
For the quark mass  $m_q = 0.35$ GeV the anomalous magnetic moment is $\mu_q \approx  0.5$.
We will also use the approximation that $\mcf_D^g\approx \mcf_D^\gamma  \approx 1 $ for Dirac form factors in both the
quark-photon and quark-gluon vertices.
In recent papers by Roberts and collaborators  \cite{Roberts:2007ji,Chang:2010hb} the anomalous magnetic
and anomalous chromomagnetic moments of the light quarks were calculated
 within the Dyson-Schwinger Equations (DSE) approach, with the non-perturbative quark and gluon propagators.
 The results are in qualitative agreement with the instanton model prediction. Unfortunately, the authors did not
 calculate both Pauli form factors at non-zero transfer momentum and, therefore, the calculation of SSA in SIDIS  within their model
 is not possible at the present time.

\begin{figure}
\begin{center}
\includegraphics*[trim=-5 0 0 0, clip,width=10.cm]{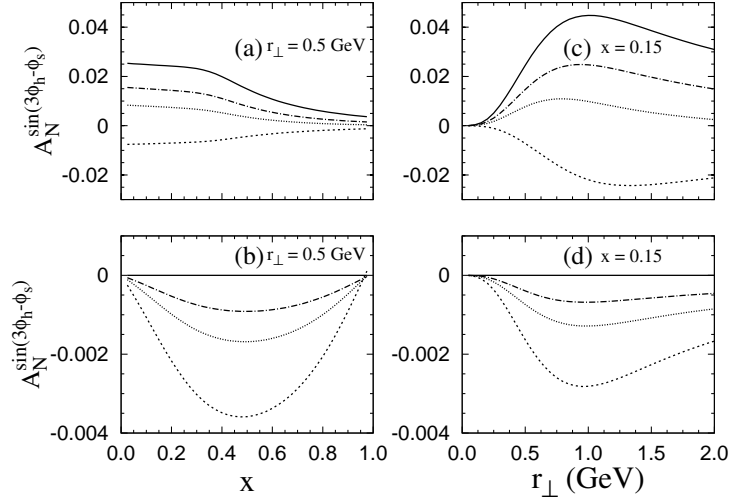}
\caption{The Collins-like asymmetry $A_N^{\sin(3\phi_h-\phi_s)}$ in the scalar (a,c) and axial-vector diquark (b,d) models.
The black dashed, dotted, and dash-dotted lines are the full results with $Q^2 =$ 1.0 GeV$^2$, 3.0 GeV$^2$ and 6.0 GeV$^2$ respectively.
The solid curves are the results without the Pauli coupling in the photon-guark vertex (${\mcf_P^\gamma} = 0$).
\label{fig:pretzelosity}}
\end{center}
\end{figure}

\begin{figure}
\begin{center}
{\includegraphics*[width=10.cm]{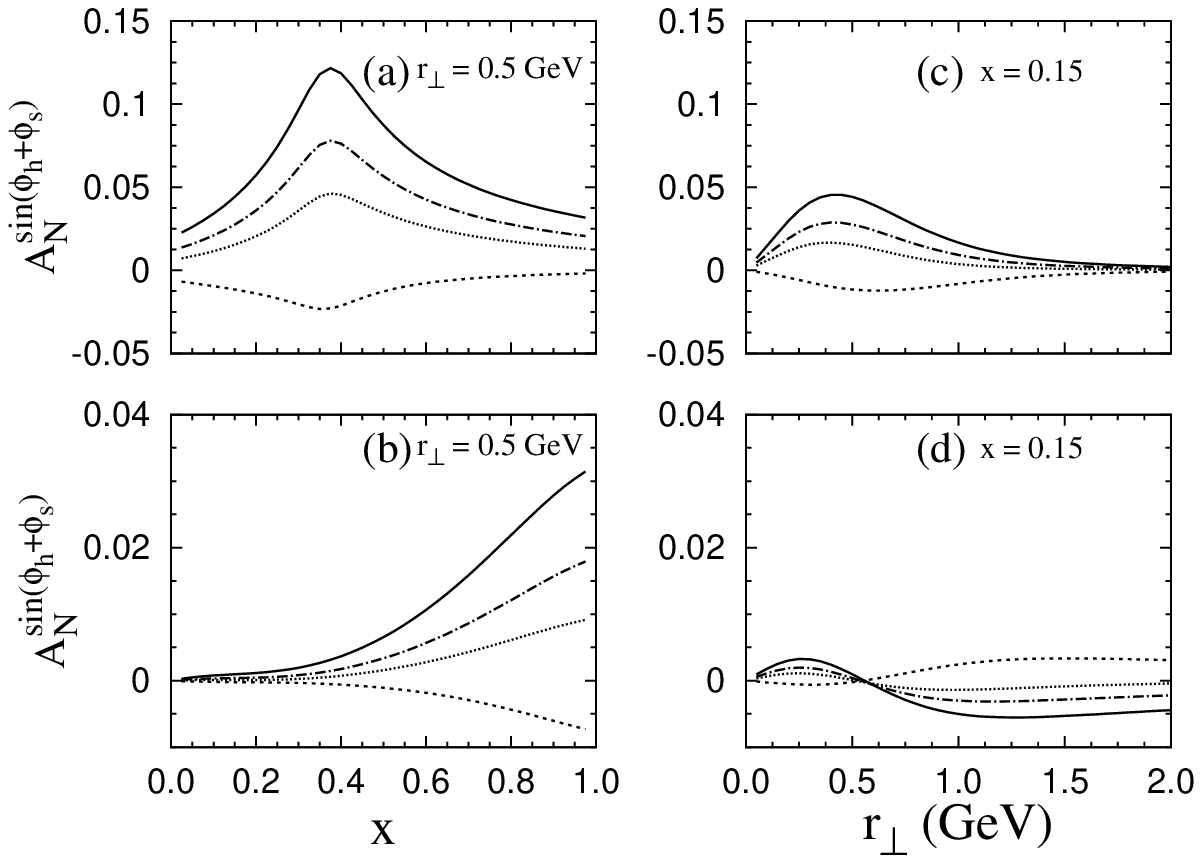}}
\caption{The  Collins-like asymmetry $A_N^{\sin(\phi_h+\phi_s)}$ in the scalar (a,c) and axial-vector diquark (b,d) models.
The notation is the same as in FIG.~\ref{fig:pretzelosity}.
\label{fig:collins}}
\end{center}
\end{figure}
\begin{figure}
\begin{center}
{\includegraphics*[width=10.cm]{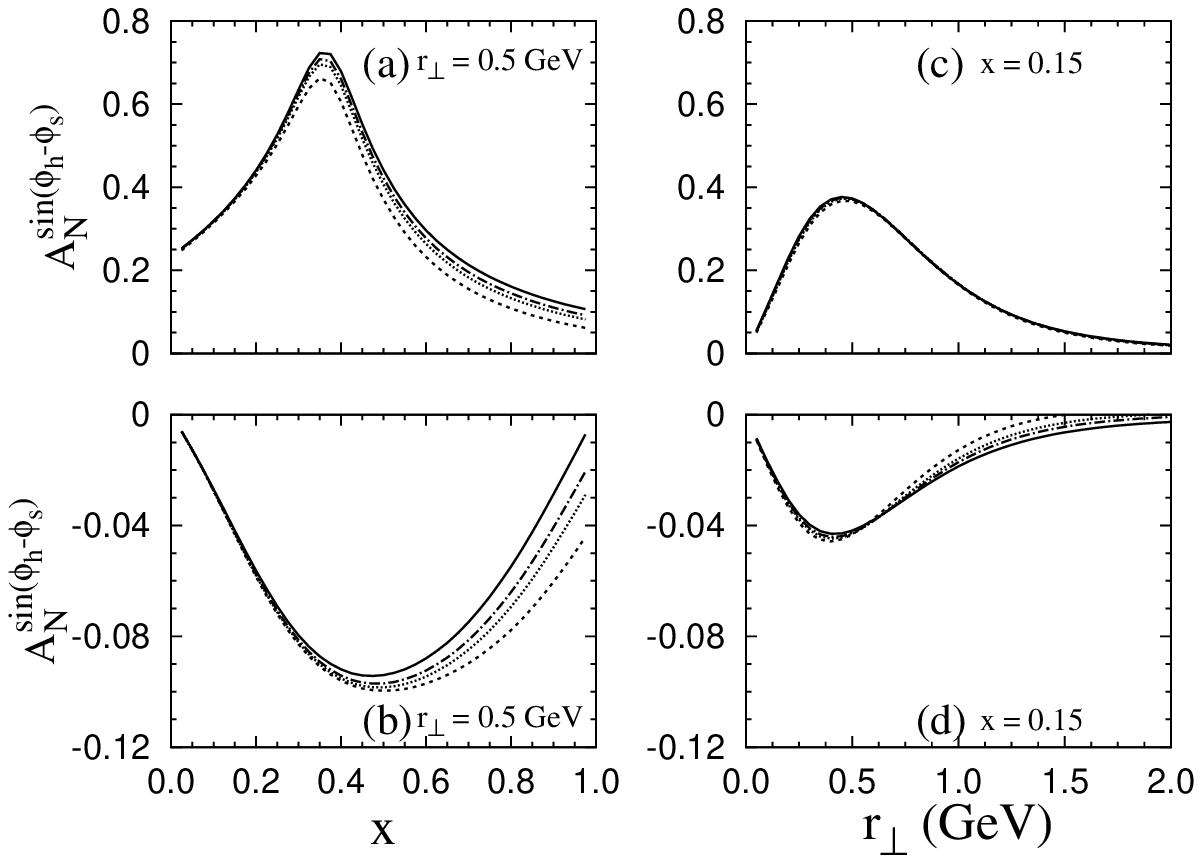}}
\caption{The Sivers-like asymmetry $A_N^{\sin(\phi_h-\phi_s)}$ in the scalar (a,c) and axial-vector diquark (b,d) models.
The notation is the same as in FIG.~\ref{fig:pretzelosity}.
\label{fig:sivers}}
\end{center}
\end{figure}
\section{Numerical results} \label{sec:numer}

We fix the value of $e_{s,a}e_q= 4 \pi C_F \alpha_s$ with $C_F = 4/3$ and $\alpha_s = 0.5$ following the arguments
in Refs.~\cite{Brodsky:2002cx,Bacchetta:2003rz,Bacchetta:2008af}. Additionally, we use $m_D = 0.6$~GeV for the mass of the scalar diquark
and $m_D = 0.8$~GeV for the axial-vector diquark. The value of the quark mass $m_q = 0.35$ GeV  is chosen  as
in the conventional diquark models~\cite{Brodsky:2002cx,Bacchetta:2003rz,Bacchetta:2008af}. This value is in agreement with
the prediction for the dynamical quark mass within the Diakonov-Petrov instanton liquid model~\cite{Diakonov:2002fq}.
Finally, the physical nucleon mass  $M = 0.94$~GeV is used.
The numerical results for asymmetries $A_N^{\sin(3\phi_h-\phi_s)}$, $A_N^{\sin(\phi_h+\phi_s)}$ and $A_N^{\sin(\phi_h-\phi_s)}$
in both the spectator models for three different values
of the $Q^2$  are presented in FIG.~\ref{fig:pretzelosity}, FIG.~\ref{fig:collins} and FIG.~\ref{fig:sivers}. The SSAs are shown as a function of Bjorken's variable $x$ with the value of transverse momentum of struck quark
$ r_\perp = 0.5$ GeV in panels (a) and (b) of those figures, and as a function of $ r_\perp$ at $x = 0.15$ in panels (c) and (d). The magnitudes of all three asymmetries
in the axial-vector diquark model are smaller than those in the scalar diquark model. However, it can be seen that in most cases
the induced Collins-like $A_N^{\sin(3\phi_h-\phi_s)}$ and $A_N^{\sin(\phi_h+\phi_s)}$ asymmetries are still of considerable size and the Pauli coupling
in the quark-photon vertex contributes to both the magnitude and shape of the asymmetries. However, for the Sivers-like $A_N^{\sin(\phi_h-\phi_s)}$
asymmetry, the effect of $\mcf_P^\gamma$ in the quark-photon interaction is small and as a result its $Q^2$ dependence is very weak.

\section{Conclusion} \label{sec:conclusion}

We have investigated the contribution  from the Pauli couplings  in both the quark-gluon and quark-photon vertices to the SSA in SIDIS,
adopting the scalar and axial-vector diquark models for the nucleon.
The specific angular dependence of SSA induced by these couplings is the same as those usually called Collins and Sivers asymmetries.
Our results show that these contributions are significant, especially for the Collins-like asymmetry.
The important observation is that it is not only the helicity flip term from the quark-gluon interaction, considered by  Hoyer and Jarvinen \cite{Hoyer:2005ev},
contributes to the angular dependence of the asymmetry $A_N$. The helicity flip from the non-perturbative quark-photon vertex is also important.
In this connection, we would like to stress that for the case of the axial-vector diquark, the Collins-like  $A_N^{\sin(3\phi_h-\phi_s)}$ asymmetry
is present only if the Pauli coupling of the photon with the struck quark is non-zero. Our results show that the
Collins-like asymmetries have significant $Q^2$ dependence for  both of the models. This effect is related to the strong $Q^2$ dependency of the quark electromagnetic Pauli
form factor in Eq.~(\ref{eq:pauliq}).

Our estimations of SSA in SIDIS are based on
the instanton model for the non-perturbative QCD vacuum.
The instantons, nonperturbative fluctuations of the vacuum gluon fields, describe
the non-trivial topological structure of the QCD vacuum and give a natural explanation
of many of fundamental phenomena of the strong interaction, such as the spontaneous chiral symmetry breaking (SCSB).
The average size of the instantons $\rho_c\approx 1/3$~fm is
much smaller than the confinement size $R_c\approx 1$~fm and can be considered as the scale of SCSB.
Furthermore, SCSB induced by the instantons is responsible for the formation of the constituent
massive quark with size $R_q\approx \rho_c\approx 1/3$~fm.  As a result, it leads to a helicity flip in both the
quark-gluon and quark-photon vertices.
Therefore, the study of the SSAs in SIDIS  might give important information on the mechanism of SCSB in the strong interaction.
Unfortunately, direct comparison of our results with experimental data on the SSA for the inclusive meson production in SIDIS
is not possible at present. It requires to introduce a quark fragmentation into account. However, our results might be relevant
to the SSA in jets production in SIDIS, which can be studied at future Electron-Ion Colliders.

\appendix*

\section{Abbreviated functions in SSA}

The abbreviated notations $\mcf^n_{s,a}$ and $J_{s,a}^n$ in Eq.~(\ref{eq:ANscalc}) are explicitly expressed as:

\bea
\mcf^+_{s} J_{s}^+ &=& \int \frac{d^2 \vec{k}_{\bot}}{(2\pi)^2} \left[ \mcf^+_{s,1} J_{s,1}^+ + \mcf^+_{s,2} J_{s,2}^+ \right]{\mathbf ,}
\nonumber \\
\mcf^-_{s} J^-_{s} &=& \int \frac{d^2 \vec{k}_{\bot}}{(2\pi)^2} \left[ \mcf^-_{s,1} J^-_{s,1} + \mcf^-_{s,2} J^-_{s,2} \right]{\mathbf ,}
\nonumber\\
\mcf^0_{s} J^0_{s} &=& \int \frac{d^2 \vec{k}_{\bot}}{(2\pi)^2} \left[ \mcf^0_{s,1} J^0_{s,1} + \mcf^0_{s,2} J^0_{s,2}
+ \mcf^0_{s,3} J^0_{s,3} -\mcf^0_{s,4} J^0_{s,4} \right] {\mathbf ,}
\nonumber\\
\mcf^+_{a} J_{a}^+ &=& \int \frac{d^2 \vec{k}_{\bot}}{(2\pi)^2} \left[ \mcf^+_{a,1} J_{a,1}^+ - \mcf^+_{a,2} J_{a,2}^+
- \mcf^+_{a,3} J_{a,3}^+ \right] {\mathbf ,} \\
\mcf^-_{a} J^-_{a} &=& - \int \frac{d^2 \vec{k}_{\bot}}{(2\pi)^2} \left[ \mcf^-_{a,1} J^-_{a,1} + \mcf^-_{a,2} J^-_{a,2}
+ \mcf^-_{a,3} J^-_{a,3} + \mcf^-_{a,4} J^-_{a,4} \right] {\mathbf ,} \nonumber\\
\mcf^0_{a} J^0_{a} &=& \int \frac{d^2 \vec{k}_{\bot}}{(2\pi)^2} \Big[ - \mcf^0_{a,1} J^0_{a,1} - \mcf^0_{a,2} J^0_{a,2}
+ \mcf^0_{a,3} J^0_{a,3} + \mcf^0_{a,4} J^0_{a,4} \nonumber \\ && - \mcf^0_{a,5} J^0_{a,5}
+ \mcf^0_{a,6} J^0_{a,6} + \mcf^0_{a,7} J^0_{a,7} + \mcf^0_{a,8} J^0_{a,8} \Big].
\nonumber
\eea

The definitions of functions $\mcf^n_{i,j}$ in the scalar diquark model are
\bea
\mcf^+_{s,1} &=& \lf(\mcf_D^\gamma - \frac{\mcf_P^\gamma}{2m_q} D_Q \rg) \lf[ \mcf_D^g
\frac{\mcf_P^\gamma}{2m_q} + \frac{\mcf_P^g}{2m_q} \lf(\mcf_D^\gamma - \frac{\mcf_P^\gamma}{2m_q} D_Q \rg)\rg], \nonumber \\
\mcf^+_{s,2} &=& \lf(\frac{\mcf_P^\gamma}{2m_q} \rg)^2 \frac{\mcf_P^g}{2m_q} = \mcf^-_{s,2} , \nonumber\\
\mcf^-_{s,1} &=& \lf(\mcf_D^\gamma + \frac{\mcf_P^\gamma}{2m_q} D_R \rg) \lf[ \mcf_D^g \frac{\mcf_P^\gamma}{2m_q} + \frac{\mcf_P^g}{2m_q}
\lf(\mcf_D^\gamma + \frac{\mcf_P^\gamma}{2m_q} D_R \rg)\rg] , \nonumber\\
\mcf^0_{s,1} &=& \lf(\mcf_D^\gamma + \frac{\mcf_P^\gamma}{2m_q} D_R \rg) \lf(\mcf_D^\gamma - \frac{\mcf_P^\gamma}{2m_q} D_Q \rg)
\mcf_D^g , \\
\mcf^0_{s,2} &=& \lf(\frac{\mcf_P^\gamma}{2m_q} \rg)^2 \mcf_D^g , \nonumber\\
\mcf^0_{s,3} &=& \lf(\mcf_D^\gamma - \frac{\mcf_P^\gamma}{2m_q} D_Q \rg)\, \frac{\mcf_P^\gamma}{2m_q}\, \frac{\mcf_P^g}{2m_q} , \nonumber\\
\mcf^0_{s,4} &=& \lf(\mcf_D^\gamma + \frac{\mcf_P^\gamma}{2m_q} D_R \rg) \frac{\mcf_P^\gamma}{2m_q} \frac{\mcf_P^g}{2m_q} . \nonumber
\eea
 Note that the loop functions $J^{\pm,0}_{s,a}$ are  real. Substituting $\vec{k}_{\bot} \to \vec{k}_{\bot}+\vec{r}_{\bot}$
after the transformation $\phi \to \phi + \psi$, we find expressions
\bea
J^+_{s,1} &=& \frac{J^0_{s,2}}{D_R} =  \frac{\left(k_\perp e^{i\phi}+r_\perp \right)\,k_\perp e^{+i\phi}\,r_\perp }{ K }
F^s_{\rm di}({\vec{k}_\perp^2}){\mathbf ,}
\nonumber\\
J^+_{s,2} &=& \frac{\left(k_\perp e^{i\phi}+r_\perp \right)^2\,k_\perp e^{-i\phi}\,r_\perp^2 }{K} F^s_{\rm di}({\vec{k}_\perp^2}){\mathbf ,}
\nonumber \\
J^-_{s,1} &=& D_R\,J^0_{s,1} = \frac{D_R^2\,k_\perp e^{-i\phi} }{K} F^s_{\rm di}({\vec{k}_\perp^2}){\mathbf ,}
\nonumber \\
J^-_{s,2} &=& \frac{D_R^2\,\left(k_\perp e^{-i\phi}+r_\perp \right)\,k_\perp e^{+i\phi}\,r_\perp}{K} F^s_{\rm di}({\vec{k}_\perp^2}){\mathbf ,}
\\
J^0_{s,3} &=& \frac{2\,D_R\,k_\perp e^{i\phi}\lf( k_\perp \cos\phi + r_\perp  \rg)\,r_\perp}{K} F^s_{\rm di}({\vec{k}_\perp^2}){\mathbf ,}
\nonumber \\
J^0_{s,4} &=& \frac{D_R\,k_\perp e^{-i\phi} [(k_\perp e^{i\phi}+r_\perp )^2+ r_\perp^2 ]}{K} F^s_{\rm di}({\vec{k}_\perp^2}){\mathbf ,}
\nonumber
\eea
with $K = \left(\vec{k}_\perp^2+\vec{r}_\perp^2+2 k_\perp r_\perp \cos\phi +B_R^2(m_q^2)\right)\vec{k}_\perp^2$.
For the axial-vector diquark model we define the combinations of couplings
\bea
\mcf^+_{a,1} &=& \mcf_D^g \lf( \frac{\mcf_P^\gamma}{2m_q} \rg)^2 \frac{x}{1-x} = \mcf^0_{s,2} \frac{x}{1-x} ,
\nonumber \\
\mcf^+_{a,2} &=& \frac{\mcf_P^g}{2m_q} \frac{\mcf_P^\gamma}{2m_q} \lf( \mcf_D^\gamma + \frac{\mcf_P^\gamma}{2m_q} D_R \rg) \frac{x}{1-x}
= \mcf^0_{s,4} \frac{x}{1-x} ,
\nonumber \\
\mcf^+_{a,3} &=& \frac{\mcf_P^g}{2m_q} \lf( \frac{\mcf_P^\gamma}{2m_q} \rg)^2 \lf(\frac{x}{1-x} \rg)^2
= \mcf^+_{s,2} \lf(\frac{x}{1-x} \rg)^2 ,
\nonumber \\
\mcf^-_{a,1} &=& \frac{\mcf_P^g}{2m_q} \lf(\mcf_D^\gamma + \mcf_P^\gamma \rg) \lf[ \mcf_D^\gamma
+ \frac{\mcf_P^\gamma}{2m_q} (x\,D_R - D_Q) \rg] \frac{x}{(1-x)^2} ,
\nonumber \\
\mcf^-_{a,2} &=& \frac{\mcf_P^g}{2m_q} \lf( \frac{\mcf_P^\gamma}{2m_q} \rg)^2 \frac{x}{(1-x)^2}
= \mcf^+_{s,2} \frac{x}{(1-x)^2} ,
\nonumber \\
\mcf^-_{a,3} &=& \mcf_D^g \frac{\mcf_P^\gamma}{2m_q} \lf(\mcf_D^\gamma + \mcf_P^\gamma \rg) \frac{x}{(1-x)^2} ,
\nonumber \\
\mcf^-_{a,4} &=& \mcf_D^g \frac{\mcf_P^\gamma}{2m_q} \lf[ \mcf_D^\gamma  + \frac{\mcf_P^\gamma}{2m_q} (x\,D_R - D_Q) \rg]
\frac{x}{(1-x)^2} ,
 \\
\mcf^0_{a,1} &=& \mcf_D^g \lf(\mcf_D^\gamma + \mcf_P^\gamma \rg) \lf( \mcf_D^\gamma + \frac{\mcf_P^\gamma}{2m_q} D_R \rg) \frac{x}{1-x} ,
\nonumber\\
\mcf^0_{a,2} &=& \mcf_D^g \lf( \frac{\mcf_P^\gamma}{2m_q} \rg)^2 \frac{x}{1-x} = \mcf^+_{a,1}{\mathbf ,}
\nonumber\\
\mcf^0_{a,3} &=& \mcf_D^g \frac{\mcf_P^\gamma}{2m_q} \lf[ \mcf_D^\gamma  + \frac{\mcf_P^\gamma}{2m_q} (x\,D_R - D_Q) \rg] \frac{x}{(1-x)^2}
= \mcf^-_{a,4} ,
\nonumber\\
\mcf^0_{a,4} &=& \mcf_D^g \frac{\mcf_P^\gamma}{2m_q} \lf(\mcf_D^\gamma + \mcf_P^\gamma \rg) \lf(\frac{x}{1-x} \rg)^2 = x\,\mcf^-_{a,3} ,
\nonumber \\
\mcf^0_{a,5} &=& \frac{\mcf_P^g}{2m_q} \frac{\mcf_P^\gamma}{2m_q} \lf(\mcf_D^\gamma + \mcf_P^\gamma \rg) \frac{x}{1-x} ,
\nonumber \\
\mcf^0_{a,6} &=& \frac{\mcf_P^g}{2m_q} \frac{\mcf_P^\gamma}{2m_q} \lf( \mcf_D^\gamma + \frac{\mcf_P^\gamma}{2m_q} D_R \rg) \frac{x}{1-x}
= \mcf^+_{a,2} ,
\nonumber \\
\mcf^0_{a,7} &=& \mcf^+_{a,3} = x\, \mcf^0_{a,8}  ,
\nonumber
\eea
and the corresponding integrals are
\bea
J_{a,1}^+ &=&
D_R\,J^+_{s,1} = J^0_{s,2} {\mathbf ,}
\nonumber
\\
J_{a,2}^+ &=&  \frac{D_R\,k_\perp e^{i\phi}\,[(k_\perp e^{i\phi} + r_\perp)^2 + \vec{r}_\perp^2] }{K} F^a_{\rm di}({\vec{k}_\perp^2}){\mathbf ,}
\nonumber \\
J_{a,3}^+ &=&  \frac{2\,k_\perp e^{i\phi}\, (k_\perp e^{i\phi} + r_\perp) (k_\perp \cos\phi + r_\perp)\,\vec{r}_\perp^2 }{K}
F^a_{\rm di}({\vec{k}_\perp^2}){\mathbf ,}
\nonumber
\\
J_{a,1}^- &=&
\frac{J^{0*}_{s,3}}{D_R}{\mathbf ,}
\nonumber
\\
J_{a,2}^- &=& \frac{ 2\,k_\perp e^{+i\phi} \, (k_\perp e^{-i\phi} + r_\perp)\,(k_\perp \cos\phi +  r_\perp) \vec{r}_\perp^2}{K}
F^a_{\rm di}({k_\perp^2}){\mathbf ,}
 \nonumber
\\
J_{a,3}^- &=& J_{a,3}^{0*} = \frac{ ((k_\perp e^{-i\phi} + r_\perp)^2 - r_\perp\,(k_\perp e^{+i\phi} +  r_\perp)) r_\perp}{K}
F^a_{\rm di}({k_\perp^2}){\mathbf ,}
\\
J_{a,4}^- &=&
\frac{J^{-}_{s,2}}{D_R^2} = \frac{J^{0*}_{a,2}}{D_R} = J_{a,4}^{0*} = \frac{J^{0*}_{a,5}}{2\,D_R} {\mathbf ,}
\nonumber
\\
J^0_{a,1} &=& J^0_{s,1},
\nonumber
\\
J^0_{a,6} &=& \frac{D_R\,k_\perp e^{+i\phi}\,( \vec{k}_\perp^2 + 2\,r_\perp k_\perp \cos\phi + 2\,\vec{r}_\perp^2 )}{K}
F^a_{\rm di}({\vec{k}_\perp^2}) {\mathbf ,}
  \nonumber \\
J^0_{a,7} &=& \frac{2\,k_\perp e^{+i\phi}\,( \vec{k}_\perp^2 + 2\,r_\perp k_\perp \cos\phi + \vec{r}_\perp^2 )\,\vec{r}_\perp^2}{K}
F^a_{\rm di}({\vec{k}_\perp^2}) {\mathbf ,}
\nonumber
\\
J^0_{a,8} &=& \frac{2\,k_\perp e^{+i\phi}\,( \vec{k}_\perp^2 \cos2\phi + 2\,r_\perp k_\perp \cos\phi + \vec{r}_\perp^2 )\,\vec{r}_\perp^2}{K} F^a_{\rm di}({\vec{k}_\perp^2}).
\nonumber
\eea

\begin{acknowledgments}

We would like to thank Prof. Vicente Vento and Chris Halcrow for carefully reading the manuscript and enlightening suggestions.
This work was supported by the National Natural Science Foundation of China (Grants No. 11405222 and  No. 11575254), by the Chinese Academy of
Sciences visiting professorship for senior international scientists (Grant No. 2013T2J0011) and by
the Chinese Academy of
Sciences President's international fellowship initiative   (Grant No 2017PM0043).

\end{acknowledgments}

\end{document}